\documentclass[pdftex,twocolumn,epjc3]{svjour3}          

\RequirePackage[T1]{fontenc}

\smartqed  

\RequirePackage{mathptmx}      
\RequirePackage{flushend}
\RequirePackage[numbers,sort&compress]{natbib}

\usepackage{graphicx,subfig}
\usepackage{float}
\usepackage{bm}
\usepackage{amsmath}
\usepackage{amsfonts,amssymb}
\usepackage{color}
\definecolor{v}{rgb}{0.6, 0.2, 0.8} 
\usepackage{soul} 
\usepackage{enumerate}
\usepackage{float}
\usepackage{subfig}
\usepackage{tabularx}
\usepackage{multicol}        
\usepackage{bm}	
\usepackage[draft]{hyperref}

\journalname{Eur. Phys. J. C}

\begin{document}

\title{Cosmological constraints on alternative model to Chaplygin fluid revisited}


\author{A. Hern\'andez-Almada\thanksref{e1,addr1}
        \and
        Juan Maga\~na\thanksref{e2,addr2}
        \and
        Miguel A. Garc\'{\i}a-Aspeitia\thanksref{e3,addr3, addr4}
        \and
        V. Motta\thanksref{e4,addr2}
}

\thankstext{e1}{e-mail: ahalmada@uaq.mx}
\thankstext{e2}{e-mail: juan.magana@uv.cl}
\thankstext{e3}{e-mail: aspeitia@fisica.uaz.edu.mx}
\thankstext{e4}{e-mail: veronica.motta@uv.cl}

\institute{Facultad de Ingenier\'ia, Universidad Aut\'onoma de Quer\'etaro, Centro Universitario Cerro de las Campanas, 76010, Santiago de Quer\'etaro, M\'exico \label{addr1}
          \and
          Instituto de F\'isica y Astronom\'ia, Facultad de Ciencias, Universidad de Valpara\'iso, Avda. Gran Breta\~na 1111, Valpara\'iso, Chile.\label{addr2}
          \and
          Unidad Acad\'emica de F\'isica, Universidad Aut\'onoma de Zacatecas, Calzada Solidaridad esquina con Paseo a la Bufa S/N C.P. 98060, Zacatecas, M\'exico \label{addr3}
          \and
          Consejo Nacional de Ciencia y Tecnolog\'ia, Av. Insurgentes Sur 1582. Colonia Cr\'edito Constructor, Del. Benito Ju\'arez C.P. 03940, Ciudad de M\'exico, M\'exico. \label{addr4}
}

\date{Received: date / Accepted: date}

\maketitle

\begin{abstract}
In this work we explore an alternative phenomenological model to Chaplygin gas proposed by H. Hova et. al. \cite{Hova2017}, consisting on a modification of a perfect fluid, to explain the dynamics of dark matter and dark energy at cosmological scales immerse in a flat or curved universe. Adopting properties similar to a Chaplygin gas, the proposed model is a mixture of dark matter and dark energy components parameterized by only one free parameter denoted as $\mu$. We focus on contrasting this model with the most recent cosmological observations of Type Ia Supernovae and Hubble parameter measurements. Our joint analysis yields a value  $\mu = 0.843^{+0.014}_{-0.015}\,$ ($0.822^{+0.022}_{-0.024}$) for a flat (curved) universe. Furthermore, with these constraints we also estimate the deceleration parameter today $q_0=-0.67 \pm 0.02\,(-0.51\pm 0.07)$, the acceleration-deceleration transition redshift $z_t=0.57\pm 0.04\, (0.50 \pm 0.06)$, and the universe age $t_A = 13.108^{+0.270}_{-0.260}\,\times (12.314^{+0.590}_{-0.430})\,$Gyrs. We also report a best value of $\Omega_k = 0.183^{+0.073}_{-0.079}$ consistent at $3\sigma$ with the one reported by Planck Collaboration.
Our analysis confirm the results by Hova et al, this Chaplygin gas-like is a plausible alternative to explain the nature of the dark sector of the universe.
\end{abstract}



\section{Introduction} \label{Int}

The accelerated expansion of the Universe is an evidence provided by type Ia supernovae (SNIa) \cite{Riess:1998,Perlmutter:1999}, and the large-scale structure (LSS) \cite{Abbott:2017wau}, being one of the major challenges in modern Cosmology. Currently, the model preferred by these observations is the so-called $\Lambda$-Cold Dark Matter ($\Lambda$CDM), which considers the dark sector as two main components: dark matter (DM) and dark energy (DE). While DM is modeled as a dust fluid and it is responsible for the large-scale structure formation in the Universe, the latter causes the accelerated expansion at late times due to its negative pressure commonly associated with the cosmological constant ($\Lambda$).

Due to $\Lambda$CDM tensions \cite{Buchert:2016}, particularly those associated with the problems that unfold from the cosmological constant (CC) \cite{Weinberg,Zeldovich}, such as the well-known fine-tuning and coincidence problems \cite{amendola_tsujikawa_2010}, a plethora of models  have emerged to explain the dark energy nature as an alternative to the CC \cite{Peebles:2003}. Examples of them are quintessence \cite{quintessence:1988}, k-essence \cite{Armendariz:2001}, braneworlds \cite{Aspeitia.cteBrane:2018,GarciaAspeitia:2011xv,Acuna-Cardenas:2017pmu,Garcia-Aspeitia:2018fvw}, among others (see for instance \cite{Bamba:2012}). However, many models are focused only on a fluid (or topology) with the capability of accelerating the universe (DE), being the DM treated as a separate entity, lacking a unifying framework with the DE.

As we mention previously, while most of these models postulate two dark components, the Chaplygin gas \cite{Chaplygin} and the generalized Chaplygin-like gas (GCG) \cite{Kamenshchik:2001,Bilic:2001,Fabris:2001} propose an unique dark fluid to describe the dynamics of the dark sector with an equation of state $p=-A\rho^{-n}$ \cite{Bento2003} where $0<n\leq 1$ and $A$ is defined as positive\footnote{Notice that it is straightforward to recover the classical Chaplygin gas model when $n=1$.}. In the inflationary model context, another elliptic Chaplygin generalization has been proposed through the Hamilton-Jacobi formalism \cite{Villanueva:2015ypa}. Another studies consider modified GCG interacting with a other fluids/fields (see for instance \cite{Wu:2007st,Chakraborty:2010ze,Biswas:2018} and references therein). The interest of this kind of models lies on some of their features: 1) they show a transition from decelerated to accelerated expansion of the universe, 2) they present the simplest corrections to $\Lambda$CDM, and 3) their microscopic origins arising from string theory and braneworlds \cite{Sundrum:1999,Ogawa:2000}.

Several studies have constrained the (generalized) Chaplygin gas models in the cosmological context using SNIa, Hubble parameter measurements, gamma ray bursts, cosmic microwave background radiation and other probes, for instance see \cite{Avelino:2003,Gong:2005,Lazkoz:2012,Marttens:2017}.  
On the other hand, studies based on the matter power spectrum without baryons effects \cite{Jarosik:2007,Sandvik:2004} rule out the GCG model. However, when the LSS analysis includes a baryon component, the GCG reproduces the 2dF mass power spectrum \cite{Beca:2003}. In addition, some authors consider models including an extra DM component to the GCG to obtain a suitable mass power spectrum \cite{Bento:2004,Multamaki:2004,amendola_tsujikawa_2010}. Motivated by these results, we aim to revisit a GCG alternative model that, while keeping its advantages, could also alleviate its known weaknesses. We will test this model with the latest cosmological observations.

In this work we investigate an extension of a recent phenomenological proposal \cite{Hova2017} to a non-flat geometry; where a generalized perfect fluid model that follows the Chaplygin gas-like scheme is studied as an alternative to GCG. 
The phenomenological model supposes a mixture of unclustered DE and DM driving the late cosmic acceleration, with stable sub-horizon fluctuations, conservation of the scale invariance instead of an unified dark sector context, resolving naturally the degeneracy problem \cite{HuEisenstein:1999, Kunz} and without future finite-time cosmological singularities.
Considering the good agreement between the  CC and the  observational data at present times, this model modifies the Equation of State (EoS) of the CC by adding an extra term which is a function of the energy density of the fluid at present. Thus, at high redshifts the fluid behaves as DM i.e. the EoS tends to zero, and at the present, it behaves as DE with an EoS similar to minus one. Moreover, this kind of evolving EoS can support the evidence of DE evolution found by \cite{Zhao:2017cud}.
The strong point of this phenomenological model is its effectiveness at reproducing the Universe dynamics without requiring additional components (or free parameters) as previous/other GCG models. 
Therefore, the free parameter of the theory is constrained by the observational Hubble data (OHD) from differential age (DA) technique \cite{Jimenez:2001gg} and the joint-light-analysis (JLA) sample of SNIa \cite{Betoule:2014}.

The papers is organized as follows. In Sec. \ref{sec:theo} we state the theoretical framework of the model presented in Ref. \cite{Hova2017}. Section \ref{sec:data} provides a description of the dataset and methods used to constrain the parameters of the Chaplygin gas-like model. In Sec. \ref{sec:Res} we discuss the results obtained and finally, in Sec. \ref{sec:Con} the remarks and conclusions are presented.

\section{Theoretical Background} \label{sec:theo}

The traditional form to obtain the Chaplygin gas is through the scalar field Lagrangian written in the form:
\begin{equation}
S[g,\phi]=-\int dtd^3x\sqrt{-g}\left(\frac{1}{2}\partial_{\mu}\phi\partial^{\mu}\phi-V(\phi)\right),
\end{equation}
where $V(\phi)$ is the scalar potential usually written in the form $V(\phi)=\phi^2\ln\phi^2+V_0$ and the action is associated with the tachyonic scalar field, $\phi$, which couples with the $U(1)$ gauge field living on the world volume theory of the non-BPS brane (see \cite{Minha} for details). In the same sense, Hova and Yang \cite{Hova2017} establish the connection of the Chaplygin gas with the tachyon scalar field through the assumption of a constant potential in the form $V(\phi)\sim A^{1/2}=V_0$, where $A$ is related to the Chaplygin EoS; similarly happens for a generalized Chaplygin gas EoS (see also \cite{Hova2017}). 
Several authors \cite{Bag:2017vjp}, recently propose new tracker models involving hyperbolic scalar potentials which may give the dark energy dynamics, including the Chaplygin gas \cite{Bag:2017vjp}. 
Moreover, other authors explore the unification of dark matter, dark energy and also inflation into a single scalar field \cite{Armendariz:2001}, whose origin could come from string landscape \cite{Liddle:2006qz}. The explicit deduction of the GCG from a microscopic point of view it is not well theoretically established and the election of the generalized EoS as $\text{sinc}(\mu\pi\rho_{df0}/\rho_{df})$ in \cite{Hova2017} it is only phenomenological, therefore the respectively EoS can be written in the form $-1+\text{sinc}(\mu\pi\rho_{df0}/\rho_{df})$.
However, some clues comes from Boehmer et. al. \cite{Boehmer:2007um} where it is possible to obtain the functional form of the EoS under the assumption of a gravitational bounded Bose-Einstein condensate (BEC) as dark matter. We would expect that a similar EoS would also be important at cosmological scales.
Nevertheless, it is necessary to strengthen the study to obtain conclusively responses, which is far from the approach of the present paper. 

We start following the recipe of \cite{Hova2017}, the generalized Chaplygin gas-like EoS is expressed as 
\begin{equation}
p_{df}=-\rho_{df}+ \rho_{df} \, \text{sinc}(\mu\pi\rho_{df0}/\rho_{df})\,,
\end{equation}
being $\text{sinc}(x)\equiv\sin(x)/x$ and $\rho_{df}$ the dark fluid density, which plays the role of the mixture of DE and DM densities. In this case $\mu$ is a dimensionless parameter constrained as $\mu\gtrsim0.688$ in order to be consistent with the stellar age bound\footnote{ Hova and Yang in \cite{Hova2017}, adopt $\mu\approx0.876$ in order to have an Universe age of $t\approx13.7$Gyrs.} and $\rho_{df0}$ is the present energy density of this fluid, constrained in terms of the density parameter as $\Omega_{df0}\sim0.96$ in \cite{Hova2017}.
It behaves as a CC in the late times of the universe evolution and as DM at the matter domination epoch. The evolution of the EoS of the dark fluid is given by
\begin{equation}
\omega_{df}(z)\equiv-1+\frac{(z+1)^3\tan(\lambda)}{[(z+1)^6+\tan^2\lambda]\xi(z)},
\label{eq:eosdf}
\end{equation}
where $\xi(z)\equiv\text{arctan}[(z+1)^{-3}\tan\lambda]$ and $\lambda\equiv\mu\pi/2$. In order to explore the universe dynamics in this context, we consider a general Friedmann-Lema\^itre-Robertson-Walker (FLRW) metric including baryonic and radiation components, hence we write the Friedmann and acceleration equations as
\begin{eqnarray}
H^2&=&\frac{8\pi G}{3}\left(\rho_{df}+\sum_i\rho_i\right)-\frac{k}{a^2}, \label{1}\\
\frac{\ddot{a}}{a}&=&-\frac{4\pi G}{3}\Big[\left(3\,\text{sinc}\left(\frac{2\lambda\rho_{df0}}{\rho_{df}}\right)-2\right)\rho_{df}\nonumber\\&&+\sum_i(1+3\omega_i)\rho_i\Big],
\end{eqnarray}
where $H\equiv\dot{a}/a$ is the Hubble parameter, $k$ is the curvature parameter which depends on the universe geometry, and the index $i$ runs over baryonic and radiation components. A dimensionless Friedmann function, $E(z)=H(z)/H_{0}$, of Eq. \eqref{1} can be written in terms of the density parameters and the redshift as \cite{Hova2017}
\begin{equation}
E(z)^2=\frac{\lambda\Omega_{df0}}{\xi(z)}+\sum_i\Omega_{i0}(z+1)^{3(1+\omega_i)}+\Omega_{k}(z+1)^2, \label{Ez}
\end{equation}
here $\Omega_{df0}\equiv 8\pi G\rho_{df0}/3H_0^2$ is the density parameter associated with the Chaplygin gas-like fluid, $\Omega_{i0}$ and $\omega_{i}$ are the density parameters and the EoS for baryonic matter and radiation\footnote{We compute $\Omega_{r0}=2.469\times10^{-5}h^{-2}(1+0.2271 N_{eff})$ \citep{Komatsu:2011}, where $N_{eff}=3.04$ is the standard number of relativistic species \citep{Mangano:2002}.}, $\Omega_{k}\equiv-k/H_0^2$ is the curvature density parameter and $H_{0}=h\times 100\, \mathrm{km\,s^{-1}Mpc^{-1}}$. In addition, we have the constraint $\Omega_{df0}+\Omega_{b0}+\Omega_{r0}=1-\Omega_{k}$. The deceleration parameter, $q(z)$, is written in the form \cite{Hova2017}
\begin{eqnarray}
&&q(z)=\frac{3\xi(z)}{2\lambda\Omega_{df0}+2\xi(z)[\sum_i\Omega_{i0}(z+1)^{3(1+\omega_i)}+\Omega_{k}(z+1)^2]}\times\nonumber\\&&\Big\lbrace\frac{\lambda\Omega_{df0}(z+1)^3\tan\lambda}{\xi(z)^2[(z+1)^6+\tan^2\lambda]}\nonumber\\&&+\sum_i(1+\omega_i)\Omega_{i0}(z+1)^{3(1+\omega_i)}+\frac{2}{3}\Omega_{k}(z+1)^2\Big\rbrace-1, \label{qz}
\end{eqnarray}
where $q(z)$ is computed by the definition $q\equiv -\ddot{a}a/\dot{a}^{2}$, which written in terms of redshift and $E(z)$ results $q(z)\equiv-1+(z+1)E^{-1}(z)(dE(z)/dz)$. As a complement, we compute the jerk parameter which is dimensionless and defined as $j=\dddot{a}/aH^3$:
\begin{eqnarray}
j(z)&=& q(z)^2+\frac{(z+1)^2}{E(z)}\frac{d^2E(z)}{dz^2} \nonumber \\
    &=&q(z)^2+\frac{(z+1)^2}{2E(z)^2}\frac{d^2E(z)^2}{dz^2}\nonumber \\ 
    & & -\frac{(z+1)^2}{4E(z)^4}\left(\frac{dE(z)^2}{dz}\right)^2,
\end{eqnarray}
where $E(z)$ and $q(z)$ come from Eqs. \eqref{Ez} and \eqref{qz} respectively, and
\begin{eqnarray}
\frac{dE(z)^2}{dz}&=&-\frac{\lambda\Omega_{df0}}{\xi(z)^2}\frac{d\xi(z)}{dz}+3\sum_i\Omega_{i0}(1+\omega_i)(z+1)^{2+3\omega_i}\nonumber \\
                  & & +2\Omega_{k}(z+1), \\
\frac{d^2E(z)^2}{dz^2}&=&\frac{2\lambda\Omega_{df0}}{\xi(z)^3}\left(\frac{d\xi(z)}{dz}\right)^2-\frac{\lambda\Omega_{df0}}{\xi(z)^2}\frac{d^2\xi(z)}{dz^2}\nonumber \\
   & & +3\sum_i\Omega_{0i}(1+\omega_i)(2+3\omega_i)(z+1)^{1+3\omega_i}\nonumber\\&&+2\Omega_{k},
\end{eqnarray}
being
\begin{eqnarray}
\frac{d\xi(z)}{dz}&=&-\frac{3\tan\lambda}{(z+1)^4\left(1+\tan^2\lambda(z+1)^{-6}\right)}, \\
\frac{d^2\xi(z)}{dz^2}&=&-\frac{18\tan^3\lambda}{(z+1)^{11}\left(1+\tan^2\lambda(z+1)^{-6}\right)^2} \nonumber \\
& & +\frac{12\tan\lambda}{(z+1)^5\left(1+\tan^2\lambda(z+1)^{-6}\right)}.
\end{eqnarray}
Note that we have followed the positive sign definition of the jerk parameter as \cite{Visser:2003vq}. Commonly, this quantity provide information on the possible evolution of any DE component. Thus, if its value is $j=1$, the DE behaves as CC, otherwise it is a dynamical dark energy fluid.

In addition, from Eqs. \eqref{Ez} and \eqref{qz} it is possible to calculate an effective EoS containing the contributions of the Chaplygin gas-like and the standard fields like baryons, radiation and the curvature term  
\begin{equation}\label{eq:weff}
\omega_{eff}(z)=-\frac{1}{3}+\frac{2}{3}q(z)\left[1+\frac{\Omega_{k}(1+z)^2}{E(z)^2}\right].
\end{equation}
Finally, using the following expression
\begin{equation}\label{eq:tA}
t_A = \frac{1}{H_0}\int_0^\infty \frac{dz'}{(1+z')E(z')},
\end{equation}
we estimate the age of the universe for the Chaplygin gas-like model.

\section{Data and methodology}
\label{sec:data}

In this section we introduce the observational data and methodology used to constrain the free parameters of the Chaplygin-like model. 

\subsection{Measurements of $H(z)$ from cosmic chronometers}

Some of the current estimation of the Hubble measurements are obtained from cosmic chronometers. In the literature, a cosmic chronometer is a passive-evolving galaxy, i.e. without ongoing star formation. The difference in age (related to $H$) is obtained by considering two of these galaxies with similar metallicities and separated by a small redshift interval \cite{Jimenez:2001gg}. The data provided by the DA method are cosmological-model-independent and can be used to probe alternative cosmological models. Here, we use 
the latest OHD obtained from DA, which contains $31$ data points covering $0 < z < 1.97$, compiled by \cite{Magana:2018} and references therein.
The  chi-square for the OHD is written as
\begin{equation}
\chi_{\mathrm{OHD}}^2 = \sum_{i=1}^{31} \frac{ \left[ H(z_{i}) -H_{DA}(z_{i})\right]^2 }{\sigma_{H_i}^{2}}
+\left(\frac{H_{0}-73.24}{1.74}\right)^{2},
\end{equation}
where $H(z_{i})$ is the theoretical Hubble parameter related to Eq. \eqref{Ez},
$H_{DA}(z_{i})$ is the observational one at redshift $z_i$, and
$\sigma_{H_i}$ its uncertainty. Notice that in the chi-square formula we also consider the measurement of $H_{0}=73.24\pm1.74\,\mathrm{Kms^{-1}Mpc^{-1}}$ \cite{Riess:2016jrr} as a Gaussian prior.

\subsection{Type Ia Supernovae}

We use the JLA compilation by Ref. \cite{Betoule:2014} consisting in $740$ SNIa in the range $0.01<z<1.2$. The observational distance modulus is computed as
\begin{equation}
\mu_{obs}=m_{B}-\left(M_{B}-a\,X_{1}+b\,{C}\right),
\label{eq:muobs}
\end{equation}
where $m_B$ is the observed peak magnitude in rest-frame $B$ band, $X_{1}$ is the time stretching of the light-curve, and
$C$ is the supernovae color at maximum brightness. The $M_B$ parameter is defined as
\begin{equation}
    M_{b}=
    \begin{cases}
      M^{1}_{b}, & \text{if the host stellar mass} \, M_{*}<10^{10}M_{\odot} \\
      M^{1}_{b}+\delta_{M}, & \text{otherwise}.
    \end{cases}
  \end{equation} 
Thus, we have two free parameter, $M_{b}^1$, and $\delta_{M}$. The quantities $a$, and $b$ are nuisance parameters in the distance estimate. 
On the other hand, the theoretical distance modulus is given by $\mu_{th}=5\log_{10}(d_{L}/\,10\,pc)$,
being $d_{L}=(1+z)D_{M}$, the luminosity distance predicted by the Chaplygin-like model and $D_{M}$ is
\begin{equation}
  D_{M}(z) =
  \begin{cases}
    \frac{c}{H_{0}\sqrt{\Omega_{k}}}\mathrm{sinh}\left[\sqrt{\Omega_{k}}\int_{0}^{z}\frac{dz^{\prime}}{E(z^{\prime})}\right] & \text{for $\Omega_{k}>0$} \\
    \frac{c}{H_{0}} \int_{0}^{z}\frac{dz^{\prime}}{E(z^{\prime})}& \text{for $\Omega_{k}=0$} \\
  \frac{c}{H_{0}\sqrt{\Omega_{k}}}\mathrm{sin}\left[\sqrt{\Omega_{k}}\int_{0}^{z}\frac{dz^{\prime}}{E(z^{\prime})}\right] & \text{for $\Omega_{k}<0$}
  \end{cases}
\end{equation}
The chi-square for SNIa data can be calculated as
\begin{equation}
\chi^{2}_{\mbox{JLA}}=\mathbf{\left(\mu_{obs} - \mu_{th}\right)^{\dag}\mathrm{C_{\eta}^{-1}}\left(\mu_{obs} - \mu_{th} \right)}, \label{fJLA}
\end{equation}
where $\mathrm{C_{\eta}}$ is the covariance matrix of the measurements provided by \cite{Betoule:2014}.

\subsection{Joint analysis}

To provide stronger constraints, we also perform a joint statistical analysis by combining the OHD and SNIa datasets. The chi-square function results as
\begin{equation}
\chi^{2}_{\mathrm{Joint}}=\chi^{2}_{\mathrm{OHD}}+\chi^{2}_{\mathrm{JLA}}.
\end{equation}
In the following section, we present our results of the parameter estimation for the Chaplygin-like gas models.

\section{Results} \label{sec:Res}

We test two models: one is a flat universe and the other one has a  curvature term $\Omega_{k}\neq 0$.
To estimate the free model parameters we perform a Bayesian analysis employing an Affine-invariant Markov chain Monte Carlo (MCMC) method provided in the emcee Python module \cite{emcee:2013} for three data sets: OHD, SNIa and its joint analysis (i.e. OHD+SNIa). We consider a burn-in phase which is stopped when the converge is achieved, which is done by requesting that the Gelman-Rubin test is less than $1.07$ for all parameters \cite{Gelman:1992}. Then, we set $6000$ MCMC steps with $500$ walkers. We consider Gaussian priors for $h$ and $\Omega_{b0}h^2$ centered at $h=0.723\pm 0.017$ and $\Omega_{b0}h^2 = 0.02202\pm 0.00046$, and flat priors over $\mu$ and $\Omega_k$ in the range $0.60<\mu<1.0$ and $-1.0<\Omega_k<1.0$ respectively. The lower limit for $\mu$ is established to be consistent with bounds on the age of the universe of $t_A>11-12\,$Gyrs \cite{Hova2017}.

Table \ref{tab:par} provides the best fit values and their corresponding uncertainties at $68\%$ CL for both geometries of the universe. The different data sets estimate consistent values on the $\mu$ parameter and the chi-square values ($\chi^2_{min}$) indicate a good-fit of the data. The joint constraint, $\mu=0.843^{+0.014}_{-0.015}$, is within $2.4\sigma$ to the value chosen as initial condition by \cite{Hova2017} to obtain late cosmic acceleration.
On the other hand, our constraints on the curvature term under this Chaplygin-like cosmology are consistent, within $3\sigma$, with the estimated $\Omega_{k}=-0.052 ^{+0.049}_{-0.055}$ from the Planck measurements of the CMB temperature spectra \cite{Planck:2016}.
 
\begin{figure}
\centering
\includegraphics[scale=0.5]{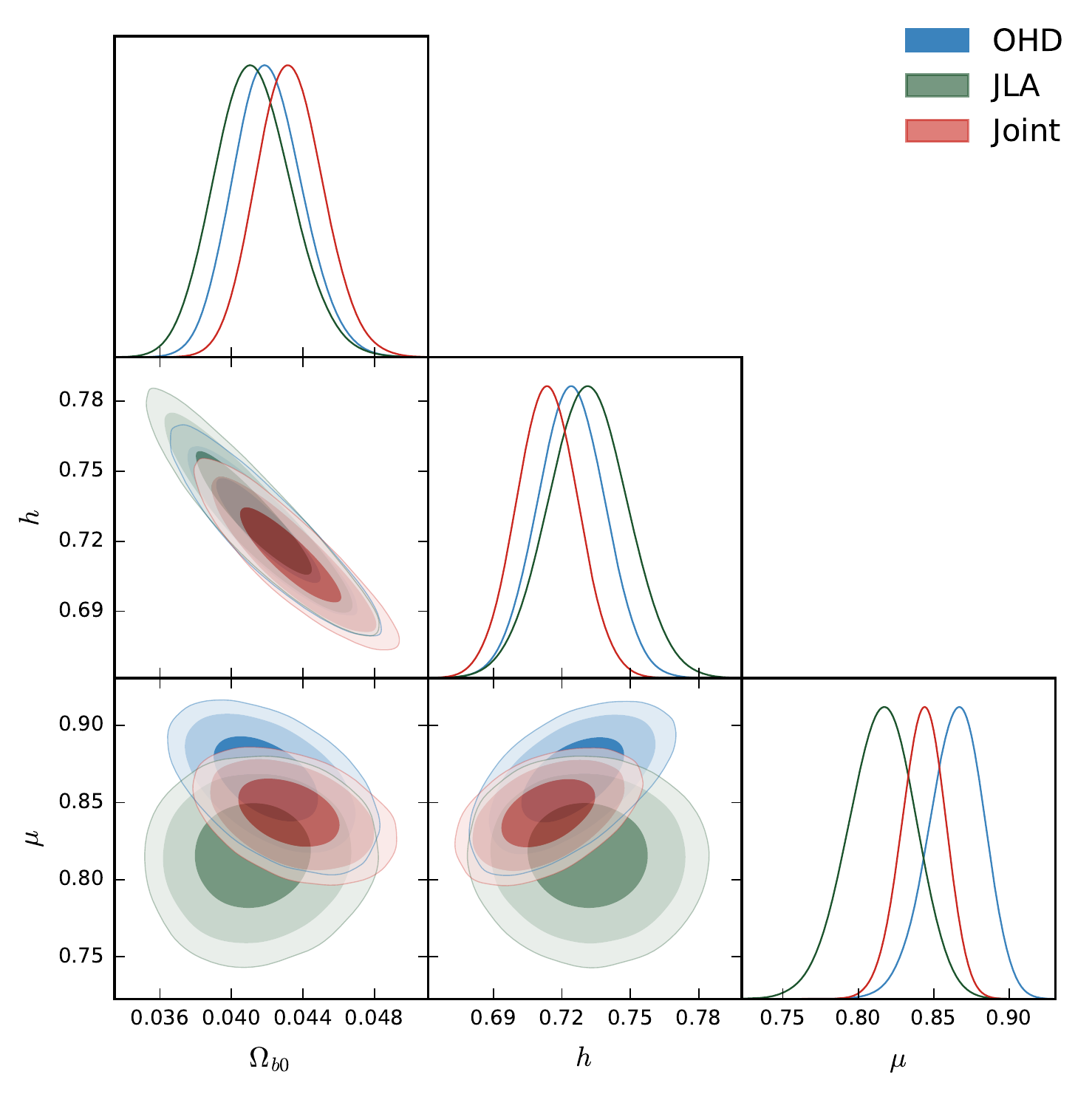}
\caption{1D marginalized posterior distributions and the 2D $68\%$, $95\%$, $99.7\%$ CL for the $\Omega_{b0}$, $h$, and $\mu$ parameters in a flat universe.}
\label{fig:contours}
\end{figure}

\begin{figure}
\centering
\includegraphics[scale=0.45]{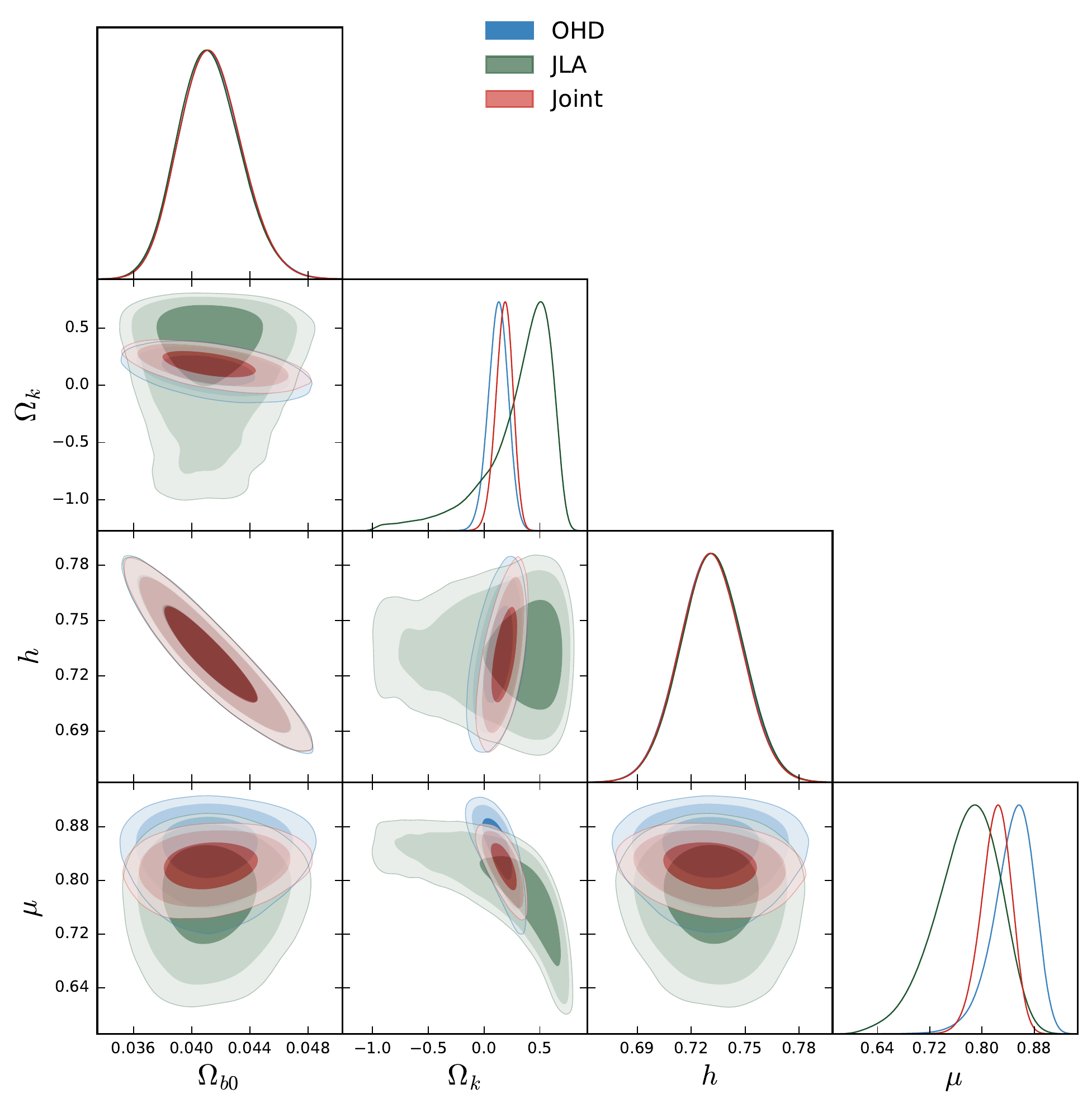}
\caption{1D marginalized posterior distributions and the 2D $68\%$, $95\%$, $99.7\%$ confidence levels for the $\Omega_{b0}$, $\Omega_{k}$, $h$, and $\mu$ parameters for a non-flat universe.}
\label{fig:contoursOK}
\end{figure}

Figure \ref{fig:contours} (Figure \ref{fig:contoursOK}) shows the 1D marginalized posterior distributions and the 2D $68\%$, $95\%$, $99.7\%$ confidence levels (CL) for the $\Omega_{b0}$, $h$, $\mu$, and ($\Omega_{k}$) parameters for a flat (curved) universe. In the flat universe, the correlations of $\mu$ with $\Omega_{b0}$ and $h$ are $\rho(\mu,\Omega_{b0})=-0.37$ and $\rho(\mu,h)=0.44$. The corresponding correlations in the non-flat universe are $\rho(\mu,\Omega_{b0})=-0.17$, $\rho(\mu,h)=-0.16$, and $\rho(\mu,\Omega_{k})=-0.70$. Following the notation in \cite{Almada-Aspeitia:2018}, the effects of $\mu$ over $\Omega_{b0}$ and $h$ are negligible when the universe is curved, but with noticeable influence over the curvature component.

\begin{figure}
\centering
\includegraphics[scale=0.45]{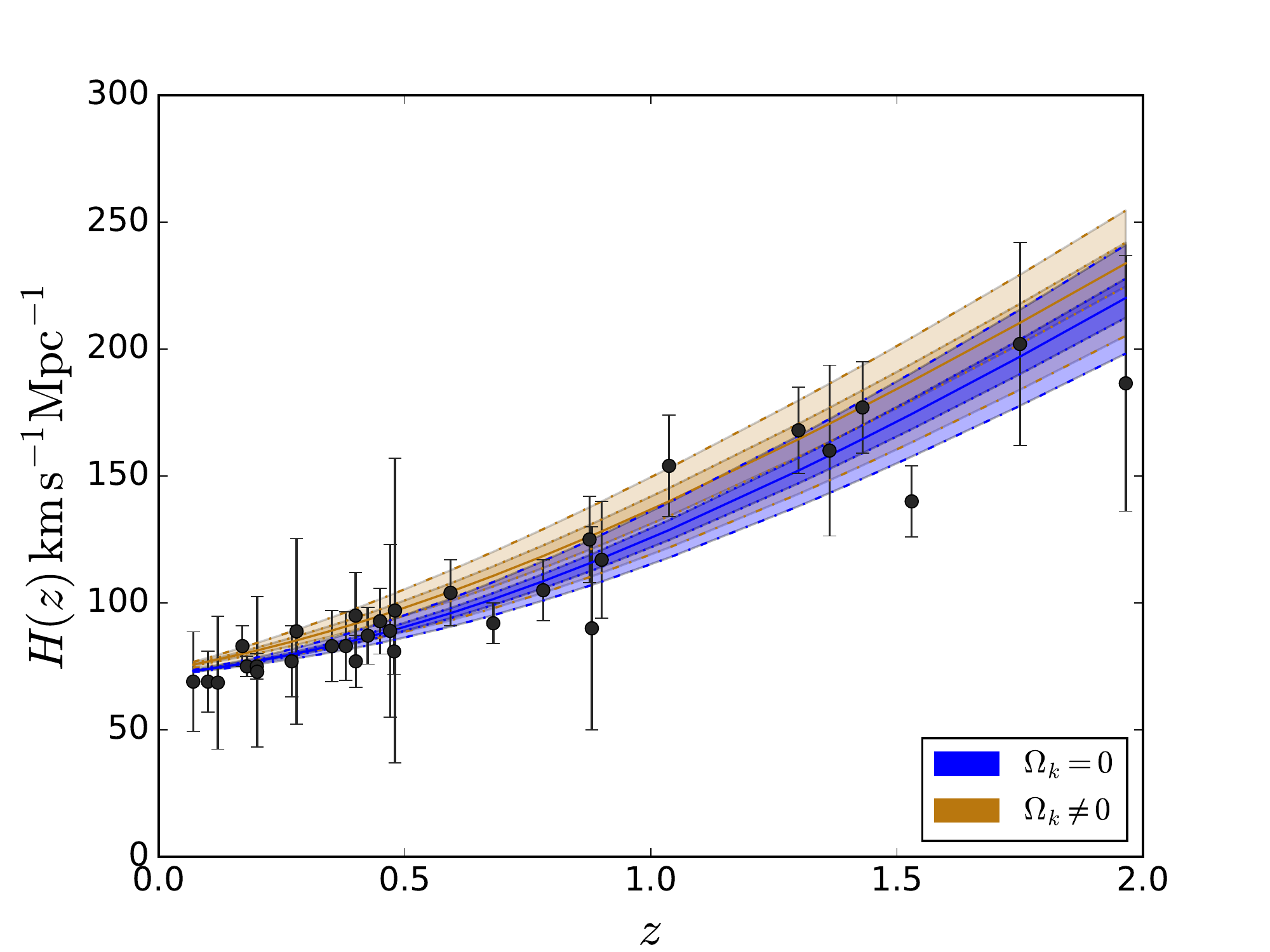}\\
\includegraphics[scale=0.45]{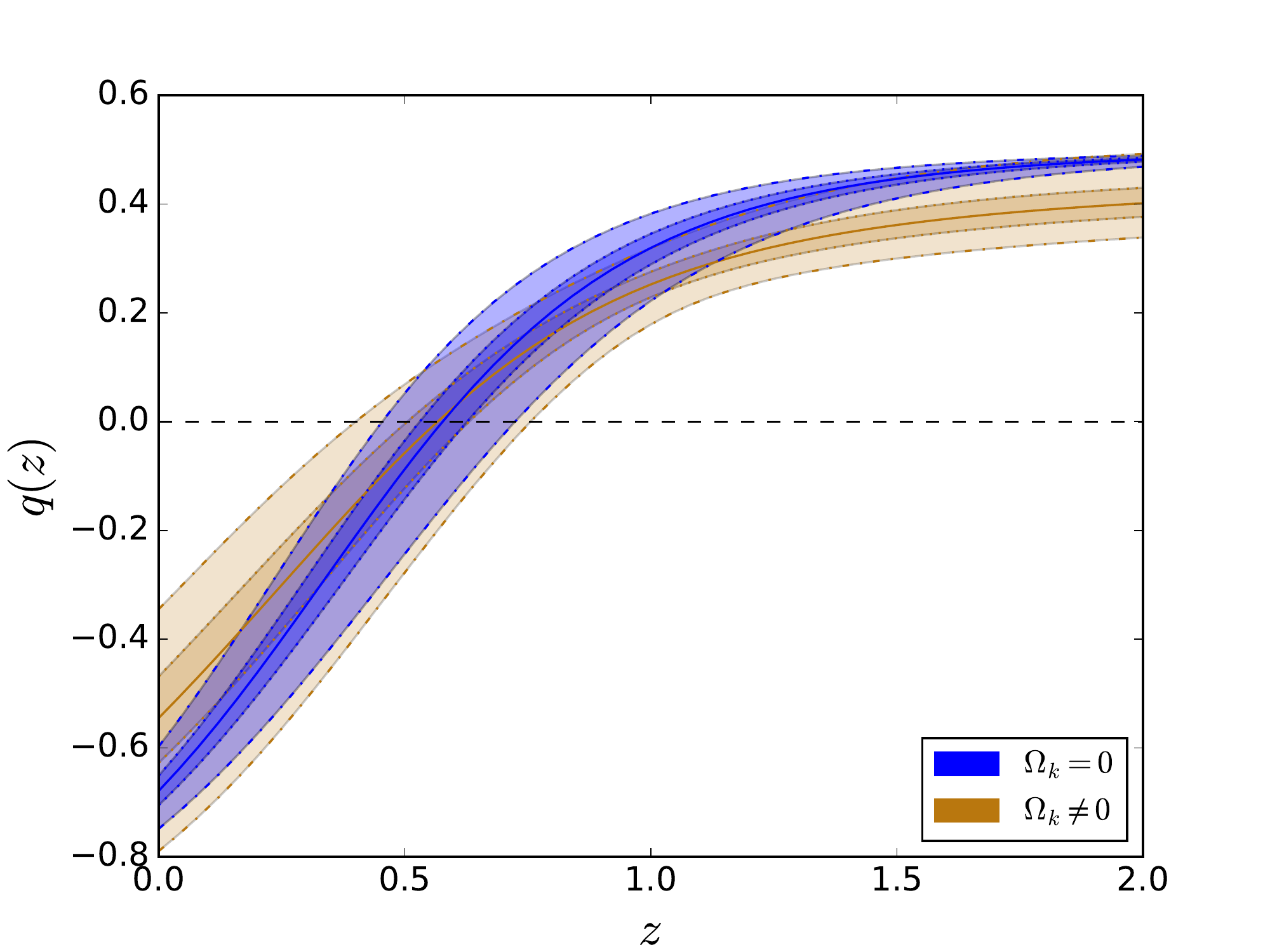}
\caption{Best fit to cosmic chronometers OHD (top panel) and reconstruction of the deceleration parameter (bottom panel) using the constraints from the joint analysis for both, flat (blue color)  and non-flat (brown color) models. The uncertainty bands refers to $68\%$ (inner band) and $99.7\%$ (outermost band) CL}
\label{fig:hzqz}
\end{figure}

\begin{figure}
\centering
\includegraphics[scale=0.45]{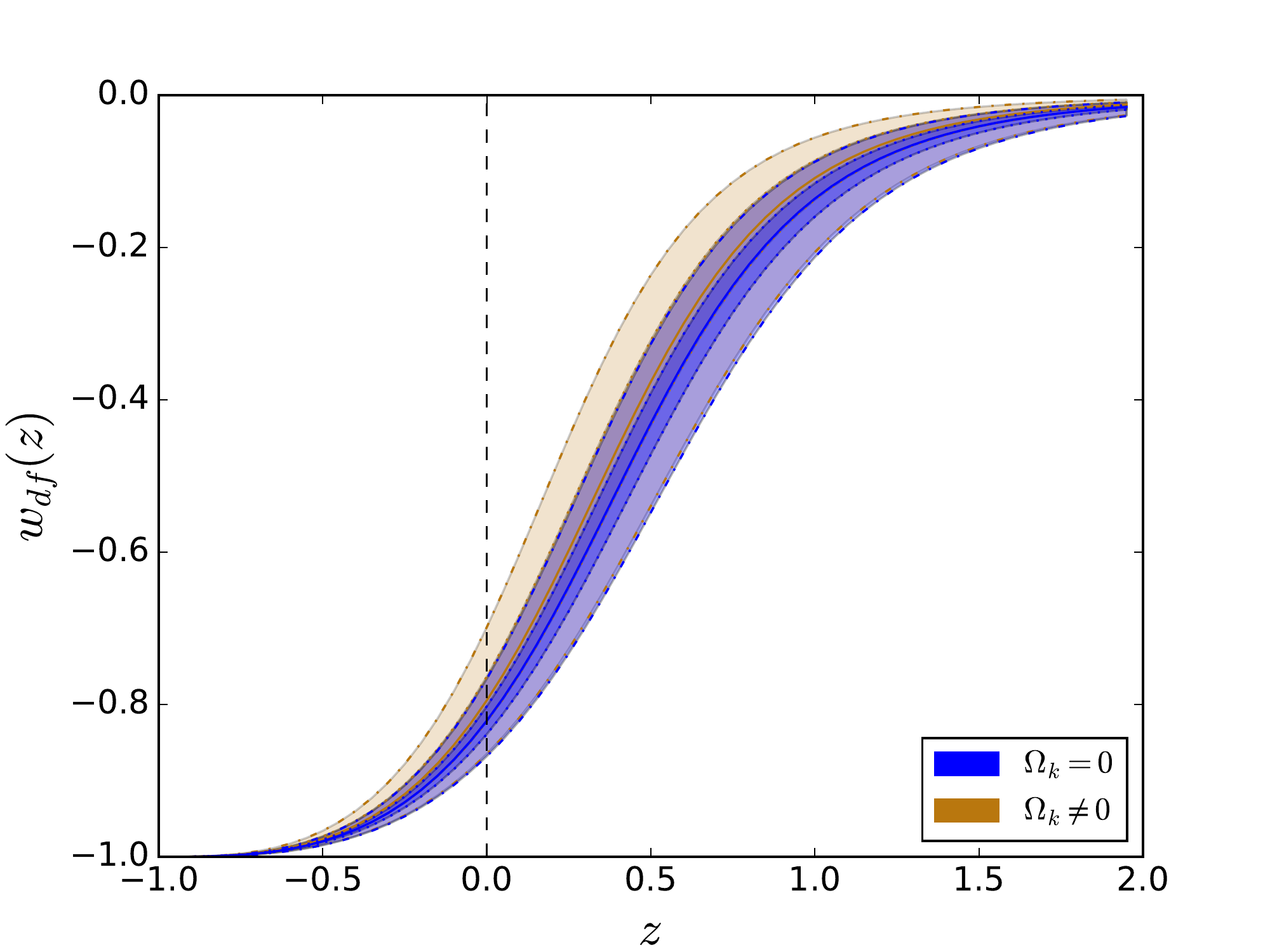}\\
\includegraphics[scale=0.45]{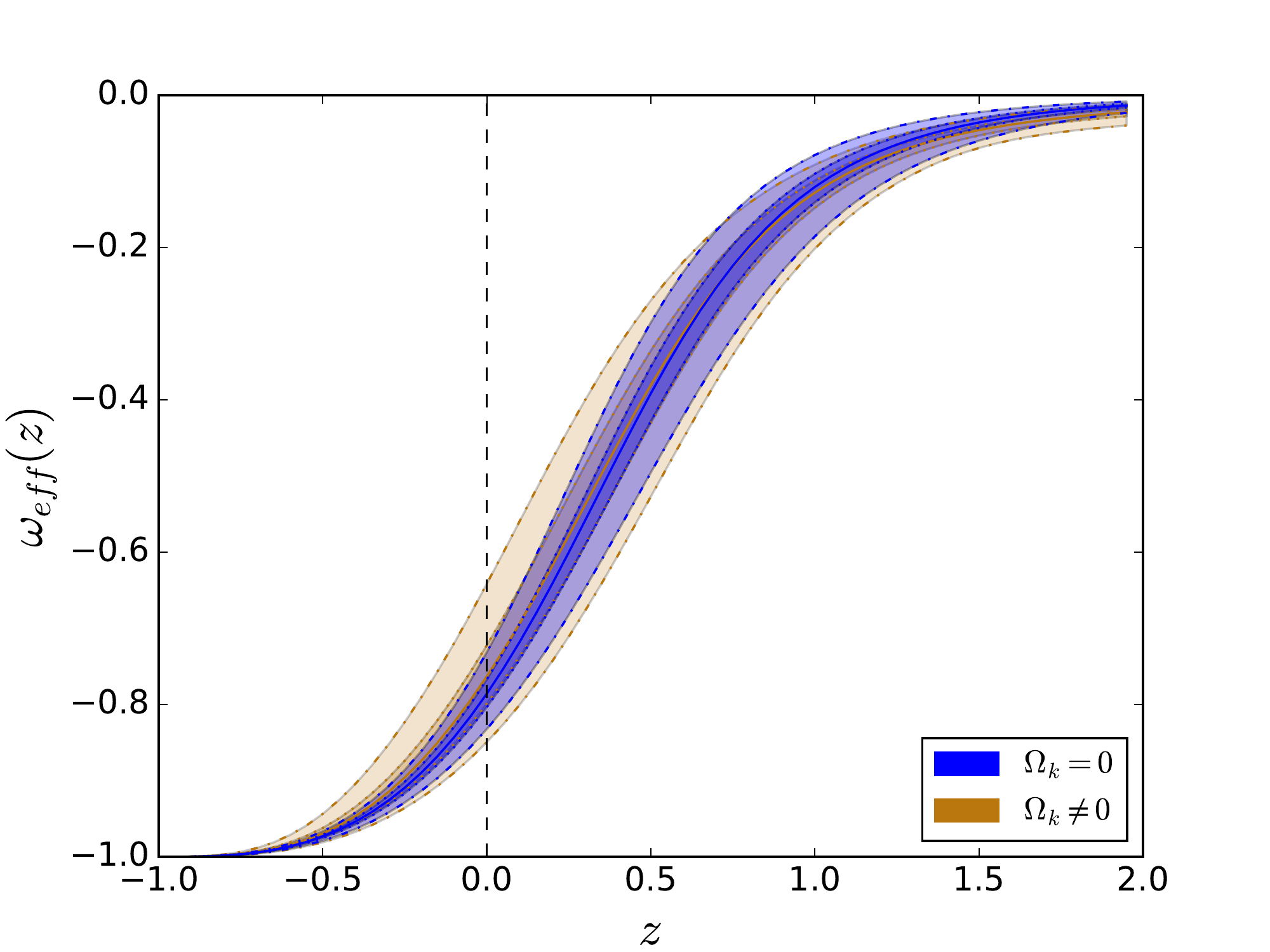}
\caption{Reconstruction of the equation of state for the dark fluid given by Eq. (\ref{eq:eosdf}) (top panel) and the effective one  (Eq. \ref{eq:weff}) (bottom panel) using the constraints from the joint analysis for both cases, flat (blue color) and non-flat (brown color) models. The uncertainty bands refers to $68\%$ (inner band) and $99.7\%$ (outermost band) CL.}
\label{fig:wz}
\end{figure}

\begin{figure}
\centering
\includegraphics[scale=0.45]{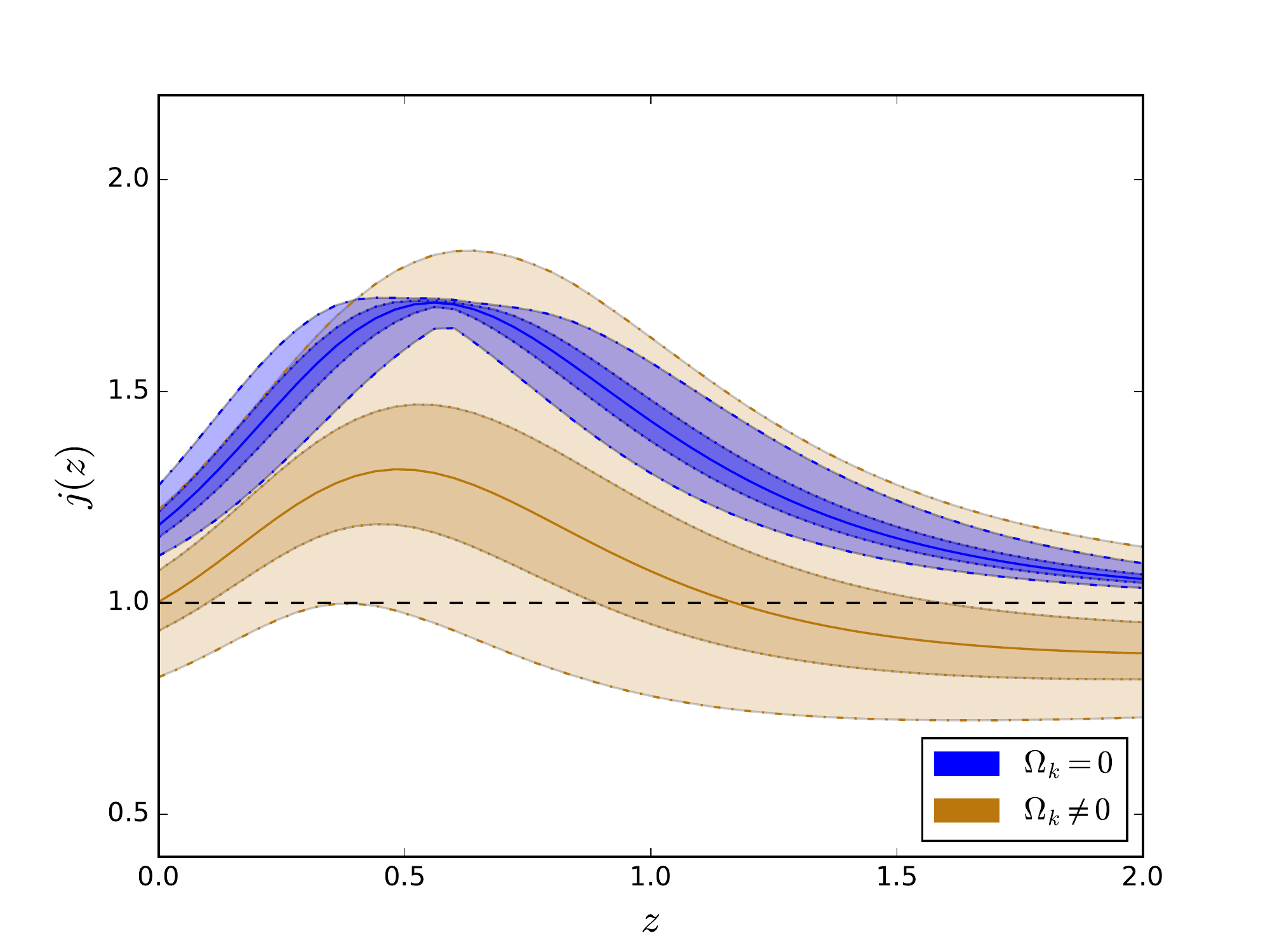}
\caption{Reconstruction of the jerk parameter using the constraints from the joint analysis for both cases, flat (blue color) and non-flat (brown color) models. The horizontal dashed line corresponds to the $\Lambda$CDM model. The uncertainty bands refers to $68\%$ (inner band) and $99.7\%$ (outermost band) CL.}
\label{fig:jz}
\end{figure}

Taking into account the best fit values of the model parameters obtained from the joint analysis, we compare the $H(z)$ and the $q(z)$ reconstruction between the spatially flat and curved universes and found that there is an agreement (within $2\sigma$) in the region $0<z<2.0$ (see Fig. \ref{fig:hzqz}). For the flat (curved) universe, the deceleration parameter at the present epoch is  $q_0 = -0.67 \pm 0.02$ ($-0.51 \pm 0.07$), which is consistent with the concordance model $q_0^{\Lambda CDM} = -0.54 \pm 0.07$, calculated from the $\Lambda$CDM mean values obtained by Ref. \cite{Garcia-Aspeitia:2018fvw}. We obtain a similar redshift, $z_{t}=0.57\pm 0.04 \, (0.50 \pm 0.06)$, for the deceleration-acceleration transition in both geometries, which is consistent within $2.5\sigma$ with  $z_{t} = 0.64^{+0.11}_{-0.06}$ 
obtained by \cite{Moresco:2016mzx} from cosmic chronometers and baryonic acoustic oscillations data for an open universe. Based on the EoS reconstruction of the dark fluid (Eq. \ref{eq:eosdf}), its behavior for both flat and non-flat cases at recent times is consistent with quintessence region and also confirms the Universe acceleration (see Fig. \ref{fig:wz}). In addition, for both models, the effective EoS (Eq. \ref{eq:weff}) at $z\gtrsim 2$ is achieved for $\omega_{eff}\to 0$, indicating that the dynamics of the universe is dominated by a non-relativistic fluid, which is consistent with our hypothesis of the Chaplygin gas-like. Moreover, the $\omega_{df}$ behavior at $z\lesssim 0.5$ also confirms that the dark fluid behaves like a quintessence field, which dominates the dynamics of the universe. On the other hand, the jerk parameter, presented in Fig. \ref{fig:jz}, shows a clear deviation, more than $3\sigma$ CL, with respect to a perfect fluid (jerk equal to one) in a flat universe; this reinforces the idea of a dynamical DE. However, for a non-flat universe the jerk parameter may mimic the perfect fluid within $3\sigma$ CL.

We estimate the universe age by using the expression \eqref{eq:tA} and the joint analysis, obtaining $t_A = 13.108^{+0.270}_{-0.260}\,$Gyrs for a flat geometry and $t_A = 12.314 ^{+0.500}_{-0.430}$ Gyrs for a curved one. The results are, as expected, in agreement with the values reported by \cite{Planck:2016}, $t_A^{\rm{Planck}} = 13.799 \pm 0.021$ Gyrs, assuming a $\Lambda$CDM model.

To statistically compare both, flat and non-flat models, the Akaike information criterion (AIC) and the Bayesian information criterion (BIC) are given in Table \ref{tab:aic_bic}. We also provide the difference with respect to the minimum value for each data set. From the joint analysis, the minimum AIC and BIC values are those for the non-flat model. Thus, if the universe is filled with a Chaplygin-like fluid instead of DM and DE, a non-flat geometry is preferred for this combination of data. However, the model in both geometries are in good agreement with the observational data used.

\begin{table*}
\centering
 \resizebox{0.95\textwidth}{!}{
\begin{tabular}{|l|ccc|ccc|}
\hline
         & \multicolumn{3}{c}{Flat universe} & \multicolumn{3}{|c|}{Non-flat universe}\\
Data set &        OHD    &    JLA &     Joint &    OHD  &  JLA &      Joint  \\
\hline
$\chi^2_{min}$ & $14.9$ & $690.8$ & $706.7$ & $14.5$ & $682.4$ & $699.3$ \\ [0.5ex]
$\Omega_{b0}$  & $0.042^{+0.002}_{-0.002}$ & $0.041^{+0.002}_{-0.002}$ & $0.043^{+0.002}_{-0.002}$ & $0.041^{+0.002}_{-0.002}$ & $0.041^{+0.002}_{-0.002}$ & $0.041^{+0.002}_{-0.002}$ \\ [0.7ex]
$\Omega_{k}$   & - & - & - & $0.128^{+0.086}_{-0.090}$ & $0.392^{+0.187}_{-0.369}$ & $0.183^{+0.073}_{-0.079}$ \\ [0.7ex]
$h$            & $0.724^{+0.015}_{-0.015}$ & $0.724^{+0.018}_{-0.017}$ & $0.714^{+0.014}_{-0.014}$ & $0.731^{+0.017}_{-0.017}$ & $0.731^{+0.017}_{-0.017}$ & $0.731^{+0.017}_{-0.017}$ \\ [0.7ex]
$\mu$          & $0.865^{+0.018}_{-0.019}$ & $0.816^{+0.021}_{-0.023}$ & $0.843^{+0.014}_{-0.015}$ & $0.850^{+0.027}_{-0.034}$ & $0.781^{+0.045}_{-0.055}$ & $0.822^{+0.022}_{-0.024}$ \\ [0.7ex]
$a$            & -& $0.141^{+0.007}_{-0.007}$& $0.142^{+0.007}_{-0.007}$ & -& $0.141^{+0.007}_{-0.007}$ & $0.142^{+0.007}_{-0.007}$ \\ [0.7ex]
$b$            & -& $3.11^{+0.08}_{-0.08}$ & $3.12^{+0.08}_{-0.08}$ & -& $3.11^{+0.08}_{-0.08}$ & $3.11^{+0.08}_{-0.08}$ \\ [0.7ex]
$M_b^1$        & -& $-19.00^{+0.06}_{-0.06}$ & $-19.01^{+0.04}_{-0.04}$ & -& $-19.418^{+0.480}_{-0.360}$ & $-19.134^{+0.072}_{-0.073}$ \\ [0.7ex]
$\delta_M$     & -& $0.07^{+0.02}_{-0.02}$ & $0.07^{+0.02}_{-0.02}$ & -& $-0.071^{+0.023}_{-0.023}$ & $-0.070^{+0.023}_{-0.023}$ \\ [0.7ex]
\hline
\end{tabular}}
\caption{Mean values for the model parameters ($\Omega_{b0}$, $\Omega_{k}$, $h$, $\mu$)
 derived from OHD and SNIa measurements for a flat universe (left side) and non-flat one 
 (right side).}
\label{tab:par}
\end{table*}

\begin{table*}
\caption{Values for AIC and BIC for each data set  for the flat and non-flat cases. The $\Delta$ refers the difference of these values with respect the minimum value.}
\centering
\resizebox{0.9\textwidth}{!}{
\begin{tabular}{|lcccccc|}
\hline
Data set & $\mathrm{AIC}_{\Omega_{k}=0}$ & $\mathrm{AIC}_{\Omega_{k}\neq0}$  & $\Delta_{\mathrm{AIC}}$ & $\mathrm{BIC}_{\Omega_{k}=0}$ & $\mathrm{BIC}_{\Omega_{k}\neq0}$&  $\Delta_{\mathrm{BIC}}$  \\
\hline
OHD  & $20.9$ & $22.5$ & $1.6$ & $25.20$ & $28.23$ &3.03\\
JLA & $704.8$ & $698.4$ & $6.4$ & $714.83$  & $735.25$ & $20.41$ \\
Joint & $720.7$ & $715.3$ & $5.4$ & $753.33$ & $752.59$& $0.73$  \\
\hline
\end{tabular}}
\label{tab:aic_bic}
\end{table*}

\section{Conclusions} \label{sec:Con}
This paper is focused on the viability of a Chaplygin gas-like fluid in a curved space-time to resemble the current Universe dynamics. Inspired in the scheme of a Chaplygin gas, i.e. a unique fluid formed with the mixing of the DM and DE components, the phenomenological model proposed by \cite{Hova2017} is a modified perfect fluid that behaves as dust in the early epochs of the universe and as DE (CC) at recent times. The strength of this  model is its ability to reproduce the Universe dynamics, without the need of a DE component of unknown nature, by adding an extra term on the perfect fluid EoS. Although this one free parameter modified EoS is phenomenological, it could comes from a scalar field dynamics. We used the latest observational Hubble data from cosmic chronometers and the type Ia SN JLA compilation to constraint the cosmological parameters under this cosmology. We showed that in a flat universe the acceleration is variable, presenting a phase change at $z_t=0.57\pm 0.04 $. The jerk parameter shows a deviation of at least of $3\sigma$ CL with respect to the CC value, also implying a dynamical DE-like behavior.

The values for the density parameter of the baryonic matter are consistent with those expected for $\Lambda$CDM. The effective EoS has a dust behavior at redshifts higher than $\sim 1.5$, acting as dark matter and behaving like a fluid that fulfills the relation $\omega<-1/3$ at redshift below $\sim 0.57$. It is worth to notice that $\omega_{eff}(0)\sim-0.8$, entering the quintessence regime. We also report an estimate of universe age of about $13.108\,$Gyrs for a flat geometry.

In the context of a curved geometry of the universe, the jerk parameter of Chaplygin gas-like fluid is consistent at $3\sigma$ with $j=1$ for CC in the region of $0<z<2$. We observe a consistent (within $1\sigma$)  behavior of the dark fluid EoS (and also of the universe) between both geometries, i.e., the dark fluid also enters to the quintessence regime about $z\sim 0.57$ and we estimate an universe age of $12.314\,$Gyrs. Our best value $\Omega_k = 0.183^{+0.073}_{-0.079}$ is  compatible within $3\sigma$ to the one reported by the Planck Collaboration.

We have confirmed that this Chaplygin gas-like model can mimic the background Universe dynamics of the standard model,  e.g. its expansion rate and current cosmic acceleration. In the linear regime, perturbations under this cosmology give similar results to those of $\Lambda$CDM, thus, a comparable large-scale structure (LSS) is expected. 
Nevertheless, differences could arise from the non-linear regime of perturbations, such as the integrated Sachs-Wolfe effect in CMB, virialization of dark halos and assembly of galaxies, etc. The exploration of these possible effects requires further perturbation analysis and numerical simulations, which is beyond the scope of this work.
Finally, our results underscore the importance of the Chaplygin-like gas model as a plausible alternative to shed light onto the DE and DM nature.

\section*{Acknowledgements}
We thank the anonymous referee for thoughtful remarks and suggestions. The authors also thank Dr. Jos\'e Villanueva for the helpful discussion on the generalized Chaplygin gas. A.H. acknowledges SNI, J.M. acknowledges support from CONICYT/FONDECYT 3160674 and MAG-A acknowledges support from SNI-M\'exico and CONACyT research fellow. A.H. and MAG-A thank Instituto Avanzado de Cosmolog\'ia (IAC) collaborations. The authors thankfully acknowledge computer resources, technical advise and support provided by Laboratorio de Matem\'atica Aplicada y C\'omputo de Alto Rendimiento del CINVESTAV-IPN (ABACUS), Proyecto CONACYT-EDOMEX-2011-C01-165873.\\

\bibliographystyle{spphys}
\bibliography{references_chaplygin.bib}

\end{document}